\documentclass[prl,aps,twocolumn,showpacs,amsmath,amssymb,superscriptaddress,floatfix]{revtex4-2}

\usepackage{graphicx}
\usepackage[export]{adjustbox}
\usepackage{dcolumn}
\usepackage{upgreek}
\usepackage{bm,dsfont}
\usepackage{amssymb,mathtools}
\usepackage{booktabs}
\usepackage{color}
\usepackage{microtype}
\usepackage[T1]{fontenc}
\usepackage{lmodern}
\usepackage[shortlabels]{enumitem}
\usepackage{changes}
\usepackage{physics}
\usepackage{derivative}
\usepackage{pgfplots}
\usepackage{tikz}
\usepackage{tikz-cd}
\usepackage{placeins}
\usepackage[colorlinks,plainpages=false,pdfpagemode=UseNone,pdfstartview=FitBH]{hyperref}
\usepackage{sidecap}

\allowdisplaybreaks
\hypersetup{
colorlinks = true,
linkcolor = [rgb]{0.87,0.18,0.22},
citecolor = [rgb]{0.0,0.60,0.32},
urlcolor = [rgb]{0.20, 0.30, 0.54}}
\definecolor{eggplant}{RGB}{180,33,147}
\makeatletter\renewcommand{\fnum@table}[1]{\tablename~\thetable.}\makeatother

\definecolor{citecolor}{rgb}{0.0,0.60,0.32}

\begin{document}

\title{Vision Transformer Neural Quantum States for Impurity Models}

\author{Xiaodong Cao}
\affiliation{Suzhou Institute for Advanced Research, University of Science and Technology of China, Suzhou 215123, China}
\affiliation{School of Artificial Intelligence and Data Science, University of Science and Technology of China, Suzhou 215123, China}

\author{Zhicheng Zhong}
\email{zczhong@ustc.edu.cn}
\affiliation{Suzhou Institute for Advanced Research, University of Science and Technology of China, Suzhou 215123, China}
\affiliation{School of Artificial Intelligence and Data Science, University of Science and Technology of China, Suzhou 215123, China}

\author{Yi Lu}
\email{yilu@nju.edu.cn}
\affiliation{National Laboratory of Solid State Microstructures and Department of Physics, Nanjing University, Nanjing 210093, China}
\affiliation{Collaborative Innovation Center of Advanced Microstructures, Nanjing University, Nanjing 210093, China}

\date{\today}
\begin{abstract}
Transformer neural networks, known for their ability to recognize complex patterns in high-dimensional data, offer a promising framework for capturing many-body correlations in quantum systems. We employ an adapted Vision Transformer (ViT) architecture to model quantum impurity models, optimizing it with a subspace expansion scheme that surpasses conventional variational Monte Carlo in both accuracy and efficiency. Benchmarks against matrix product states in single- and three-orbital Anderson impurity models show that these ViT-based neural quantum states achieve comparable or superior accuracy with significantly fewer variational parameters. We further extend our approach to compute dynamical quantities by constructing a restricted excitation space that effectively captures relevant physical processes, yielding accurate core-level X-ray absorption spectra. These findings highlight the potential of ViT-based neural quantum states for accurate and efficient modeling of quantum impurity models.

\end{abstract}
\maketitle

Variational representations of quantum states offer a promising strategy for studying the rich phase diagrams of quantum many-body systems, encompassing a variety of emergent phenomena~\cite{Stewart1984,Hewson1997,Imada1998,Lee2006,Keimer2015}. While physical insights have motivated simple forms of variational wave functions that have achieved great success in explaining specific phenomena---such as the Bardeen-Cooper-Schrieffer state for superconductivity~\cite{BCS} and the Laughlin wave function for the fractional quantum Hall effect~\cite{Laughlin}---developing compact and expressive ansätze for general quantum many-body systems remains a highly nontrivial task. Recent advances in this aspect include tensor-network states~\cite{Cirac_TNreview,Orus_TNreview} and neural quantum states (NQS)~\cite{Giuseppe2017}, both of which leverage large sets of variational parameters to capture underlying entanglement and correlations. Although certain equivalences have been identified between these two categories of variational wave functions~\cite{Chen2018, NQS_tensor_Cirac,NQS_tensor_Sharir}, NQS have demonstrated superior expressive capability, particularly in encoding volume-law entanglement states~\cite{NQS_volumn_RBM, NQS_volumn_general, NQS_volumn_random, NQS_Gao}, which have remained out of reach for tensor networks~\cite{area_entanglement_Eisert}.

Despite their exceptional expressive power, NQS face significant difficulties when applied to large fermionic systems. The standard variational Monte Carlo (VMC) optimization for both the ground state and time-evolution of NQS requires extensive sampling and nontrivial regularization techniques~\cite{Giuseppe2017, NQS_criticality_Czischek, Schmitt2020} to mitigate numerical instabilities, even with autoregressive architectures~\cite{NQS_autoregressive_Donatella, NQS_RNN_Lange}. Additionally, the non-locality and intricate sign structure inherent to fermionic systems pose substantial obstacles for the application and optimization of NQS~\cite{fermion_nolocal_Nys}. Despite notable progress~\cite{NQS_fermionic_backflow_Ruggeri,NQS_fermionic_backflow, NQS_fermionic_geminal, NQS_hidden,pfau2020ferminet,paulinet,vonglehn2023psiformer,anti_han}, efficiently parameterizing ground states and accurately capturing dynamics of extensive fermionic systems with NQS remains a formidable challenge.

In this Letter, we explore this issue by studying quantum impurity models (QIMs)~\cite{anderson_imp, Kondo, Wilson_NRG} comprising up to several hundred fermionic spin-orbitals. We show that carefully designed NQS architectures and computational bases, combined with tailored optimization schemes, can effectively overcome the challenges associated with traditional VMC. Additionally, dynamical quantities and spectral functions can be computed directly on the frequency axis by selective inclusion of excited states.

\begin{figure*}[t]
\centering
  \includegraphics[width=\textwidth]{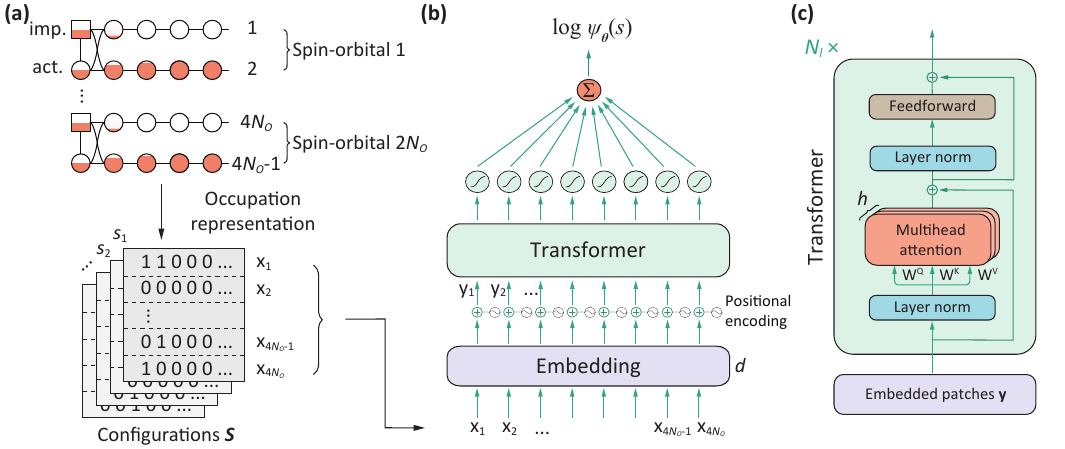}
  \caption{\label{fig::transformer} Schematic of the ViT variational wave function for a multiorbital QIM with $N_o$ orbitals. (a) The impurity model is represented in the natural orbital basis at the top (refer to the text for details). Each square represents an impurity spin-orbital, and each circle representes a bath site. The fill color represents electron occupation in each spin-orbital. Solid lines indicate hoppings between orbitals, and inter-orbital hoppings are omitted for clarity. The occupation configurations $s$ are mapped onto a binary image $\bm{x}$ and divided into $N_p=4N_o$ patches. These patches are subsequently processed through  the ViT network in (b). The details of its transformer encoder block are depicted in (c).}
\end{figure*}

\emph{Quantum impurity models.---}
The QIM we study is a multiorbital Anderson impurity model $\hat{H} = \hat{H}_\mathrm{loc} + \hat{H}_\mathrm{bath} + \hat{H}_\mathrm{hyb}$. The locally interacting impurity $\hat{H}_\mathrm{loc}$ features a rotationally invariant Kanamori form of Coulomb interaction, parameterized by the Hubbard $U$ and Hund's $J$~\cite{Georges2013}. The impurity couples to a non-interacting quantum bath $\hat{H}_\mathrm{bath}$ via hybridization $\hat{H}_\mathrm{hyb}$. We assume, without loss of generality, a bath with semi-elliptic density of states for each impurity spin-orbital, centered at zero energy with half bandwidth $D$, which serves as the energy unit throughout~\cite{SupMat}. 
We employ \emph{natural orbitals}~\cite{Loewdin1955,NO_Zgid,NO_He,NO_Lu} for the bath representation, which was shown previously to significantly simplify the entanglement structure and reduce the computational complexity in QIMs~\cite{Lu2019, cao_tree_imp_2021}. In this representation, each impurity orbital couples to a partially filled ``active'' bath orbital, which in turn couples to nearly empty or filled ``conduction'' and ``valence'' bath chains where electron or hole occupations decay exponentially into the chain [Fig.~\ref{fig::transformer}(a)]. As will be detailed later, this representation also offers further advantages in computational accuracy and efficiency within the context of NQS.

\emph{Vision Transformer NQS.---}
The ViT~\cite{Attention,ViT,NQS_transformer_Luciano_1D} wave function $\ket{\psi_{\bm{\theta}}}$ is expressed in the general form of an NQS as $\ket{\psi_{\bm{\theta}}} = \sum_{s} \psi_{\bm{\theta}}(s) \ket{s}$,
where the expansion is over a complete orthonormal computation basis set $\{\ket{s}\}$. The vector $\bm{\theta}$ represents the collection of complex-valued variational parameters in the neural network. Fermionic antisymmetry is incorporated via the Jordan-Wigner transformation~\cite{NQS_fermionic_JW}.

In the natural-orbital representation of QIMs, the computation basis $\{\ket{s}\}$ consists of strings of occupation configuration (0 or 1) for each spin-orbital, as depicted in Fig.~\ref{fig::transformer}(a). For a system with $N_o$ orbitals ($2N_o$ spin-orbitals), a configuration $\ket{s}$ can be visualized as a binary image $\mathbf{x}$ with dimensions 4$N_o$ $\times$ $N_L$, where $N_L=(N_b+1)/2$ is the length of each conduction or valence chain. To create more uniform configuration images, the conduction and valence chains are represented in electron and hole representations, respectively, where 1 indicating occupied and empty sites, respectively. Each configuration image is then divided into $N_p=4N_o$ horizontal patches $\mathbf{x}=(x_1;\cdots;x_{N_p})$. This partitioning mirrors the structure of the tree tensor-network states used in multiorbital QIMs~\cite{cao_tree_imp_2021}. 

The adapted ViT architecture, shown in Fig.~\ref{fig::transformer}(b) begins by embedding the input configuration $\mathbf{x}$ into input feature vectors $\mathbf{y}=(y_1;\cdots;y_{N_p})\in\mathbb{C}^{N_p\times d}$ via a linear projection of dimension $d$, and are subsequently added by learnable positional encoding vectors.
The key to capturing both local and global correlations within the input feature vectors is the multihead self-attention mechanism depicted in Fig.~\ref{fig::transformer}(c). Here, $\mathbf{y}$ is linearly projected onto $h$ independent sets of query ($\mathbf{q}_i = \mathbf{y} \cdot \mathbf{W}^Q_i$), key ($\mathbf{k}_i = \mathbf{y} \cdot \mathbf{W}^K_i$), and value ($\mathbf{v}_i = \mathbf{y} \cdot \mathbf{W}^V_i$) vectors ($i \in \{1,2,\dots,h\}$), where $\mathbf{W}^{Q,K,V}_i \in \mathbb{C}^{d\times d/h}$ are variational parameters. The attention vectors for the $i$-th head are computed as $\mathbf{A}_i=\text{softmax}\left(\mathbf{q}_{i}\cdot \mathbf{k}_{i}^T/\sqrt{d/h}\right)\cdot \mathbf{v}_i \in \mathbb{C}^{N_p\times d/h}$. 
The $h$ attention vectors $\mathbf{A}_i$ are then concatenated and undergo a further linear projection to integrate the representations from different heads. These mixed attention vectors are passed through a non-linearity block, consisting of a two-layer fully-connected neural network with hidden dimension $2d$ and $\text{GeLu}(\cdot)$ activation. Pre-layer normalization~\cite{PreNorm_transformer} and skip connections~\cite{resnet} are applied before and after each block. This full transformation block can be repeated $N_l$ times. The final output, representing $\log\psi_{\bm{\theta}}(s)$, is obtained by applying a $\log\left[\cosh(\cdot)\right]$ non-linearity to the transformer output, followed by a summation over all features. The hyperparameters in this network include the hidden dimension $d$, the number of attention heads $h$, and the number of transformer layers $N_l$.

\emph{Subspace expansion optimization of NQS.---}
The expectation value of an operator $\hat O$ for an given $\ket{\psi_{\bm{\theta}}}$ is
\begin{equation}\label{eq:e_loc_mc}
    O_{\bm{\theta}} = \langle \hat{O} \rangle_{\bm{\theta}} = \sum_{s} P_{\bm{\theta}}(s) O_{\bm{\theta}}^{\text{loc}}(s),
\end{equation}
where $P_{\bm{\theta}}(s) = |\langle \psi_{\bm{\theta}} | s \rangle|^2 / \langle \psi_{\bm{\theta}} | \psi_{\bm{\theta}} \rangle$ represents the probability of configuration $\ket{s}$, and the corresponding local estimator
\begin{align}\label{eq:e_loc}
    O_{\bm{\theta}}^{\text{loc}}(s) = \sum_{s'} \langle s | \hat{O} | s' \rangle \frac{\langle s' | \psi_{\bm{\theta}} \rangle}{\langle s | \psi_{\bm{\theta}} \rangle}.
\end{align}
For a Hamiltonian $\hat H$, its energy $E_{\bm{\theta}}=\langle \hat{H} \rangle_{\bm{\theta}}$ as defined in Eq.~\eqref{eq:e_loc_mc} serves as a natural loss function for ground state optimization, where Monte Carlo methods are commonly employed to estimate the required summations~\cite{Giuseppe2017}. In the natural orbital basis, the ground state of QIMs is characterized by configurations where density fluctuations are primarily at the impurity and nearby sites~\cite{Lu2019,cao_tree_imp_2021}, while deeper sites in the bath chains remain inactive. This leads to a highly skewed probability distribution $P_{\bm{\theta}}(s)$, which poses challenges for efficient sampling using traditional Monte Carlo techniques.

We propose a subspace expansion method that leverages this skewed distribution to enhance the efficiency and accuracy of expectation value estimation. As described in Eq.~\eqref{eq:expansion}, in each optimization step, we start with an initial state space $\mathcal{S}$ containing $N_s$ states. This space is expanded by adding states $s'$ from a connected space $\mathcal{S}_c$, which includes those with non-zero Hamiltonian matrix elements $\langle s | \hat{H} | s' \rangle$ for $s \in \mathcal{S}$. The expanded space $\mathcal{S}_e$ is then truncated by retaining the $N_s$ states with the highest probabilities $|\langle s |\psi_{\bm{\theta}} \rangle|^2$, updating $\mathcal{S}$ for the next iteration. 
The variational parameters, $\bm{\theta}$, are iteratively optimized by minimizing $E_{\bm{\theta}}$ according to the approximate probability distribution $\mathcal{P}_{\bm{\theta}}(s) = \sum_{s\in\mathcal{S}}\left( |\langle s | \psi_{\bm{\theta}} \rangle|^2/\sum_{s'\in\mathcal{S}} |\langle s' | \psi_{\bm{\theta}} \rangle|^2 \right)$, employing the stochastic reconfiguration (SR) scheme~\cite{VMC_SR,VMC_McMillan,Giuseppe2017}  to account for the geometry of the learning landscape~\cite{SupMat}.
\begin{equation}\label{eq:expansion}
\begin{tikzcd}
  \mathcal{S} 
  \arrow[r, 
  bend left=10, 
  "{\mathcal{S}_c=\{s': \langle s | \hat{H} | s' \rangle \neq 0,\ \ket{s}\in\mathcal{S} \}}",
  start anchor=north east,
  end anchor=north west]
  & [10em] \mathcal{S}_e = \mathcal{S}\oplus\mathcal{S}_c
  \arrow[l, 
  bend left=10, 
  "{\text{truncate w.r.t. }|\langle s | \psi_{\bm{\theta}} \rangle|^2,\ \ket{s}\in\mathcal{S}_e}",
  start anchor=south west,
  end anchor=south east]
\end{tikzcd}
\end{equation}

\emph{Restricted excitation space and spectral functions.---}
To avoid the complexities of time-domain calculations for system dynamics~\cite{Giuseppe2017,Schmitt2020,MendesSantos2023}, we propose directly computing spectral functions in the frequency domain. We focus on core-level XAS, where the creation of a core hole during excitation acts as an impurity due to strong local core-valence interaction. Initially proposed for simple metals~\cite{Anderson1967,Nozieres1969,Doniach1970}, this impurity framework is also applicable to correlated materials when the non-interacting bath effectively represents their correlated electronic structure~\cite{XAS_Gunnarsson,deGroot2008book}, a concept well developed within dynamical mean-field theory~\cite{Metzner1989,Georges1996,Kotliar2004,Kotliar2006}.

To model XAS, the impurity Hamiltonian $\hat H$ is modified to incorporate core-valence interactions at the impurity site,
\begin{equation}
    \hat H_c = \hat H - U_c \sum_\alpha \hat n_{\alpha,0} (1-\hat n_c),
\end{equation}
where $\hat n_c$ is the occupation of a core spin-orbital, and $U_c$ quantifies its interaction with the impurities orbitals. The ground state becomes the direct product of the original ground state with a filled core state $\ket{\Psi_{\bm{\theta}}^{\text{GS}}}= \ket{c} \otimes \ket{\psi_{\bm{\theta}}^{\text{GS}}}$. The XAS intensity is given as 
\begin{equation}\label{eq:xas}
    I^{\text{XAS}}_{\alpha}(\omega) = -\frac{1}{\pi}\mathrm{Im}\left[\bra{\Psi_{\bm{\theta}}^{\text{GS}}} D^\dag_{\alpha}\frac{1}{\omega + E_0 - \hat H_c + i\Gamma} D_{\alpha}\ket{\Psi_{\bm{\theta}}^{\text{GS}}}\right]
\end{equation}
where the dipole operator $D_\alpha=c^\dag_{\alpha,0}c_c$ excites a core electron into the spin-orbital $\alpha$ at the impurity. Here, $E_0$ is the ground state energy, and the inverse core-hole lifetime $\Gamma$ parameterizes decay processes not explicitly included in $\hat H_c$.

A restricted excitation space $\mathcal{S}_\text{exc} = \text{span}\{\ket{\phi_{\alpha,m}} \}$ can be constructed to capture the most relevant excited states in XAS. Starting from the initial state
\begin{equation}
    \ket{\phi_{\alpha, m=0}} = D_\alpha \ket{\Psi_{\bm{\theta}}^{\text{GS}}} = \hat{c}^\dag_{\alpha,0} \ket{\psi_{\bm{\theta}}^{\text{GS}}},
\end{equation}
we construct the space by including all possible single-particle excitations within a cluster $\mathcal{C}_{L_{\text{exc}}}$ around the impurity site with size $L_{\text{exc}}$ generated by $\hat T_{\beta, l, l'} = \hat{c}^\dagger_{\beta,l}\hat{c}_{\beta,l'}$ (where $l,l'\in \mathcal{C}_{L_{\text{exc}}},\ \beta\in [1,2N_o]$)~\cite{SupMat}:
\begin{equation}
    \ket{\phi_{\alpha,m\geq 1}} = \hat{T}_{\beta, l, l'} \ket{\phi_{\alpha, m=0}}.
\end{equation}
Here, we only consider excitations within the same spin-orbital $\beta$. The inclusion of inter-orbital and/or inter-spin excitations is necessary for cases with off-diagonal hybridization or spin-orbit coupling in $\hat H$. $\mathcal{S}_\text{exc}$ forms an effective representation of $\hat H_c$, in which the spectral functions can be directly evaluated~\cite{noorthibasis,SupMat}. The basis set size scale quadratically with $L_{\text{exc}}$, leading to a computational complexity of $\mathcal{O}(N_o^2L_{\text{exc}}^4)$ for diagonal hybridization and at most $\mathcal{O}(N_o^4L_{\text{exc}}^4)$ for general cases.

\emph{Results.---}
We benchmark ViT results against MPS for QIMs with up to three orbitals. All NQSs are optimized over 1000 steps with a linearly decreasing learning rate from $10^{-1}$ to $10^{-3}$. A linearly decreasing diagonal shift is applied for regularization~\cite{Giuseppe2017}, starting at $10^{-1}$ and reducing to $5\times 10^{-5}$. The ViT uses a constant $h=4$ heads, and $N_b=40$ bath sites for each spin-orbital are used in all cases unless otherwise specified. All MPS calculations were performed with global $U(1)_{\text{charge}} \times U(1)_{S_z}$ symmetries.

\begin{figure}[t]
\centering
  \includegraphics[width=\linewidth]{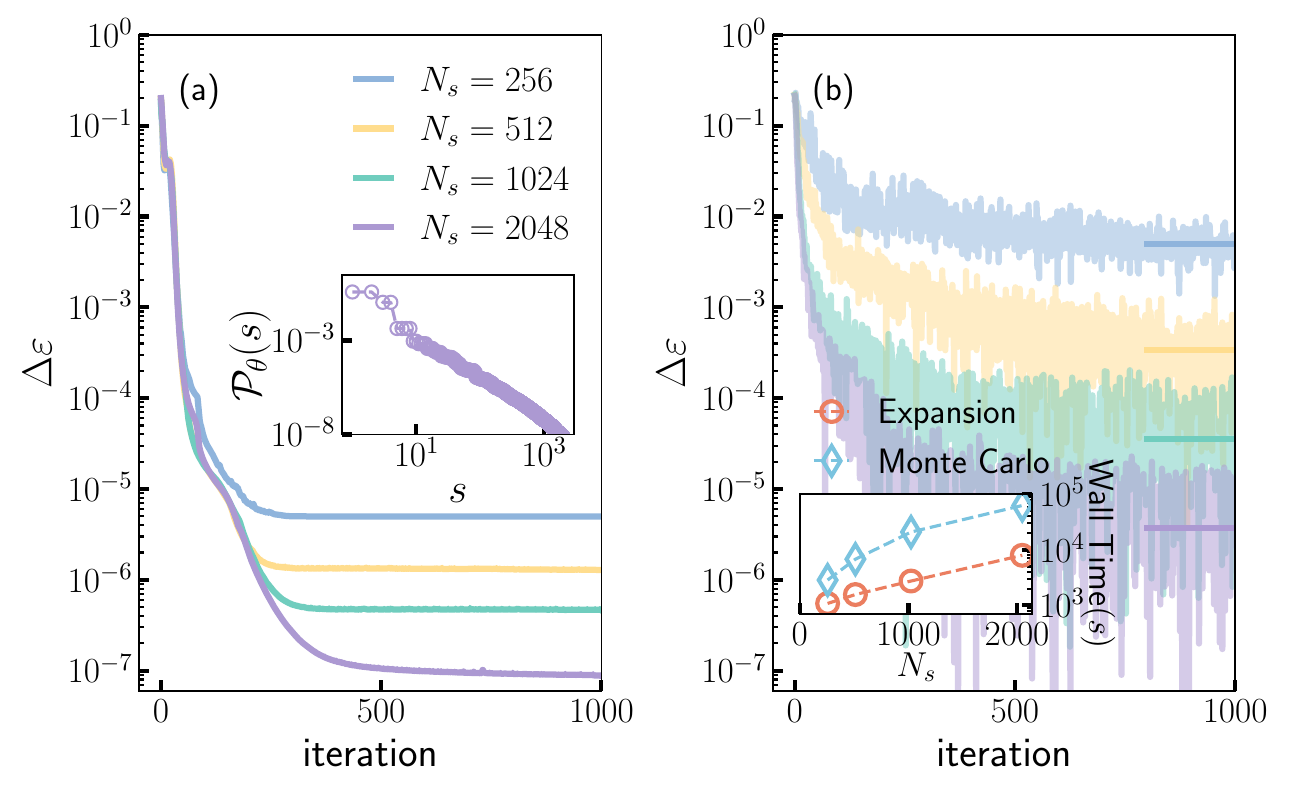}
  \caption{\label{fig::comparingMC} Relative error $\Delta\varepsilon=\left(E_{\text{ViT}} - E_{\text{DMRG}}\right)/|E_{\text{DMRG}}|$ of the single-orbital Anderson impurity model for different space sizes $N_s$, with the NQS optimized by (a) subspace expansion and (b) VMC. Horizontal lines in (b) indicate the averaged VMC energies of the last 200 iterations. The inset of (a) shows the probability weight of configurations in the optimized ground state of $N_s=2048$. The inset of (b) shows the wall time of subspace expansion optimization (red open circle) and VMC (open cyan diamond). The model parameters are $U=2D$, $\mu=-D$. We use a ViT with $N_l=1$ and $d=32$.}
\end{figure}

Figure~\ref{fig::comparingMC} compares the error in the ViT ground state energy relative to MPS, using subspace expansion in (a) and VMC (in the chain geometry~\cite{cao_tree_imp_2021}) in (b). The reference MPS ground state is computed using density matrix renormalization group (DMRG) with a bond dimension $\chi=1000$. For the same number of states $N_s$, the subspace expansion method achieves at least an order of magnitude lower error than VMC, with a much smoother training process. As expected, the enhanced accuracy is due to the skewed probability distribution $\mathcal{P}_{\bm{\theta}}(s)$ shown in the inset of Fig.~\ref{fig::comparingMC}(a). Additionally, as shown in the inset of Fig.~\ref{fig::comparingMC}(b), even with the same $N_s$, the smaller connectivity in the natural orbital basis and consequently fewer non-zero $\langle s' | \hat H | s\rangle$ elements [see Eq.~\eqref{eq:e_loc}], makes our method significantly faster, with nearly an order of magnitude speedup for $N_s\geq1024$. Notably, as shown in~\cite{SupMat}, with the same network configurations, calculations for larger systems with $N_b$ up to 100 and different $U$ values maintain consistent accuracy.

\begin{figure}[t]
\centering
  \includegraphics[width=\linewidth]{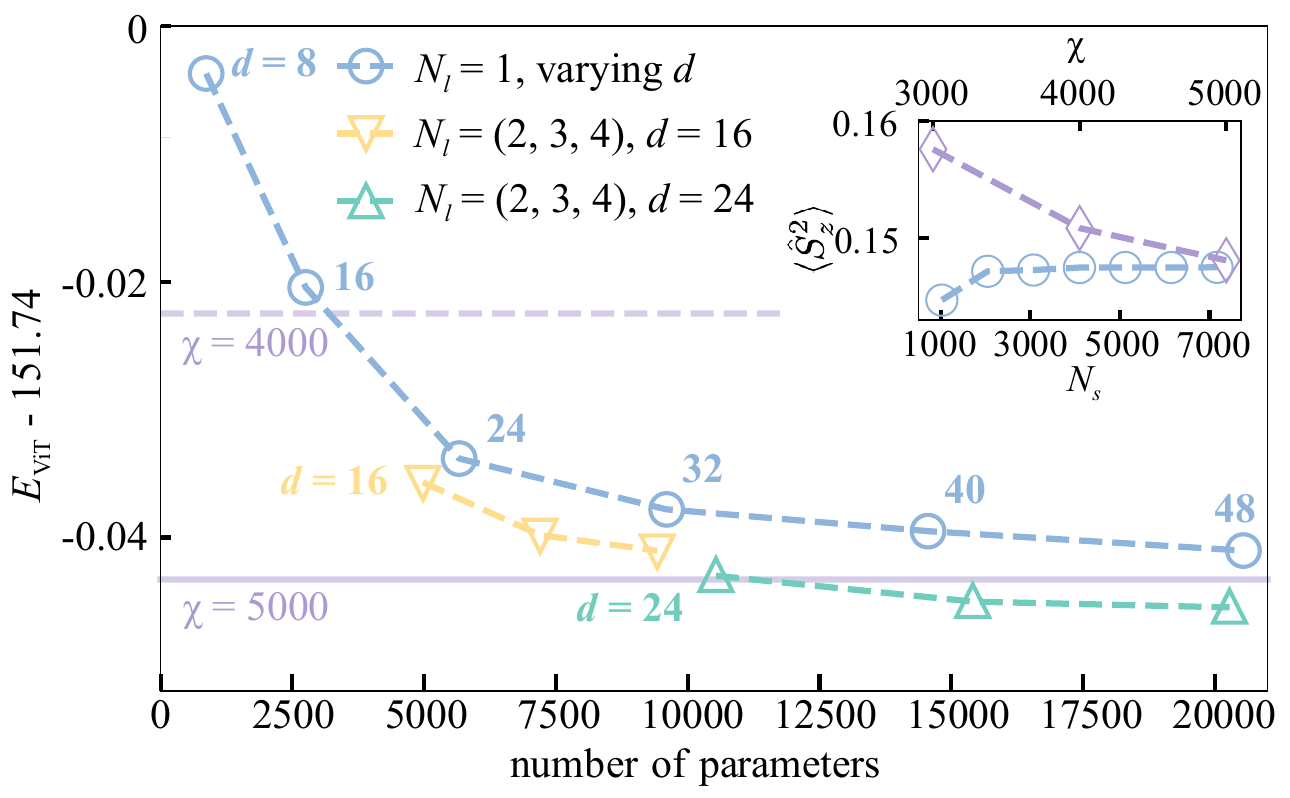}
  \caption{\label{fig::modelparameters} Ground state energy $E_{\text{ViT}}$ of the three-orbital Anderson impurity model as a function of the number of parameters of the ViT variational wave function. Parameter numbers are varied by increasing the hidden dimension $d$ from 8 to 48 with a fixed number of transformer layers $N_l=1$ (open circle), and by varying $N_l=2,3,4$ with a fixed $d=16$ (open down triangle) or $d=24$ (open up triangle). The subspace size remains constant at $N_s=4096$. Results from MPS are shown as horizontal lines for bond dimensions $\chi=4000$ (dashed) and $\chi=5000$ (solid). The inset shows local spin fluctuations $\langle \hat S_z^2 \rangle$ for different $N_s$ values for ViT with $N_l=1$ and $d=32$ (blue circle), compared to MPS results with bond dimension $\chi$ (purple diamond). The model parameters are $U=1.60D$, $J=0.25U$, and the chemical potential $\mu$ is adjusted to ensure an impurity occupation $\langle\hat{N}\rangle=2$. }
\end{figure}

Figure~\ref{fig::modelparameters} shows the performance of ViT applied to a more demanding three-orbital model case with complex entanglement due to inter-orbital correlations. Comparisons between different configurations $d$ and $N_l$ indicate that while increasing variational parameters consistently improves the ground state energy, deeper ViT architectures with larger $N_l$ generally yield better accuracy with similar parameter counts. Remarkably, a ViT with only a few thousand variational parameters (e.g., $d=16, N_l=1$) achieves ground state energies comparable to MPS results with $\chi=4000$, even though MPS utilizes significantly more parameters, scaling as $\mathcal{O}(L\chi^2)$. The inset of Fig.~\ref{fig::modelparameters} shows local spin fluctuations $\langle \hat{S}_{z}^2 \rangle$ on each impurity orbital computed with different subspace sizes $N_s$, with rapid convergence at $N_s=2048$. In contrast, MPS-DMRG results align with ViT only when $\chi$ is increased to 5000. These findings highlight the ViT’s efficiency, making it a compact yet robust wave function ansatz for multiorbital QIMs.

\begin{figure}[t]
\centering
  \includegraphics[width=\linewidth]{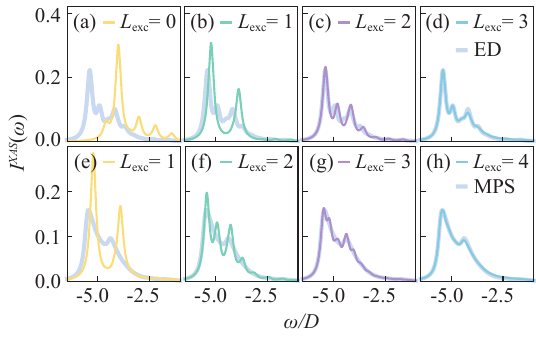}
  \caption{\label{fig::XASBenchmark} XAS for a half-filld two-orbital impurity model computed from the restricted excitation space, compared with (a)-(d) ED for $N_b=6$, and (e)-(h) MPS for $N_b=40$. The cluster size $L_\text{exc}$ increases from left to right across the panels.
  $L_\text{exc}=0$ only includes the impurity site.
  The model parameters are $U=1.60D$, $J=0.25U$, $U_c=1.25U$, and $\Gamma=0.10U$.
  }
\end{figure}

We proceed to study the dynamics of QIMs using ViT by computing the XAS of a half-filled two-orbital model within a restricted excitation space, and compare the results with ED and MPS in Fig.~\ref{fig::XASBenchmark}. The ground state is parameterized by a ViT with $N_l=1$ and $d=32$, and it is optimized using the subspace expansion method with $N_s=2048$. For MPS, the spectra are calculated in the time domain using the single-site time-dependent variational principle evolved to $T_{\text{max}}=40/D$ with a time step $dt=0.05/D$. 

The X-ray absorption involves two transitions, increasing the occupation of an impurity spin-orbital from 0 to 1 or 1 to 2. In the local limit, where bath width is neglected, the energy difference between these transitions approximates $U - U_c$, resulting in two distinct absorption peaks. In the $L_\text{exc}=0$ case, where only the impurity site is included, the two-peak feature is absent due to the lack of impurity charge fluctuations [Fig.\ref{fig::XASBenchmark}(a)]. It is recovered only when $L_\text{exc} \geq 1$, emphasizing the need for bath degrees of freedom to capture the dynamics. For $N_b=6$, the full system is included by $L_\text{exc}=3$, and the XAS becomes indistinguishable from that of ED in Fig.\ref{fig::XASBenchmark}(d), demonstrating that the restricted active space captures the essential states for XAS. Artifactual peaks appear here as the limited number of bath sites fail to reflect the continuous nature. 
Continuous spectra with typical asymmetric line shape are only recovered with a sufficiently large number of bath sites $N_b=40$. In Figs.~\ref{fig::XASBenchmark}(e)-(h), the ViT spectra rapidly converge towards MPS results. By $L_\text{exc}=4$, both the excitation peaks for empty (right peak) and singly-occupied (left peak) configurations, as well as the continuum from bath screening, are accurately reproduced, demonstrating the efficacy of the restricted excitation space in capturing the essential physics of XAS.

\emph{Conclusion and discussion.---}
We introduced an effective method for modeling QIMs using NQS based on the ViT architecture. The Transformer’s self-attention mechanism, which enables all-to-all interactions, is crucial for capturing global correlations and complex entanglements in QIMs. By incorporating the natural orbital basis, we developed a subspace expansion optimization method that achieves high-accuracy ground states with significantly reduced computational resources compared to traditional VMC methods, demonstrating superior performance in both accuracy and efficiency. Additionally, we proposed computing spectral functions by constructing a restricted excitation space that effectively captures essential physical processes, yielding accurate core-level XAS spectra. 

These findings highlight the potential of ViT-based NQS, combined with subspace expansion optimization, as a powerful tool for studying QIMs. This approach offers a promising alternative to QMC and tensor-network methods, which often struggle with sign problems or long-range entanglement in complex scenarios involving significant multiorbital or multi-impurity correlations. We anticipate that, with appropriately adapted basis optimization schemes, this approach could also be promising for quantum chemistry systems~\cite{MPS_basis_Krumnow} and lattice systems~\cite{MPS_Basis_Fishman}, where spectral functions can be calculated using specifically tailored restricted excitation spaces~\cite{Lu_RIXS_2017,Charlebois_expansion_2020}.

\begin{acknowledgments}
This work is supported by the National Key R\&D Program of China (No. 2022YFA1403000) and the National Natural Science Foundation of China (No. 12274207). All NQS results presented in this work were obtained using NetKet~3~\cite{netket3} with a single NVIDIA V100 GPU (32 GB). We thank the open-source community for their contributions to the development of JAX~\cite{jax}, Flax~\cite{flax}, and Optax~\cite{optax}, which were instrumental in producing our results.
\end{acknowledgments}

\bibliographystyle{apsrev4-2}
\bibliography{ref_clean_abbrev}

\clearpage
\onecolumngrid
\begin{center}
\large{\textbf{Supplemental Material for \\ ``Vision Transformer Neural Quantum States for Impurity Models''}}
\end{center}

\setcounter{figure}{0}
\setcounter{table}{0}
\setcounter{equation}{0}
\setcounter{page}{1}
\renewcommand{\thefigure}{S\arabic{figure}}
\renewcommand{\thetable}{S\Roman{table}}
\renewcommand{\theequation}{S\arabic{equation}}

\def\@hangfrom@section#1#2#3{\@hangfrom{#1#2}#3}
\def\@hangfroms@section#1#2{#1#2}

\section*{S1 --- Quantum Impurity Model}
The multiorbital Anderson impurity model with Kanamori interaction is defined as
\begin{align}\label{eq:anderson}
\hat{H} = &\, \hat{H}_\mathrm{loc} + \hat{H}_\mathrm{bath} + \hat{H}_\mathrm{hyb}, \\ \nonumber
\hat{H}_\mathrm{loc} = &\sum_{\{\alpha\}} \varepsilon_{\alpha_1\alpha_2}\hat{c}^\dagger_{\alpha_1,0} \hat{c}_{\alpha_2,0} \\ \nonumber
&+ U \sum_m \hat{n}_{m \uparrow,0} \hat{n}_{m \downarrow,0}+ (U-2J) \sum_{m \neq m^{\prime}} \hat{n}_{m \uparrow,0} \hat{n}_{m^{\prime} \downarrow,0}+(U-3J) \sum_{m<m^{\prime}, \sigma} \hat{n}_{m \sigma,0} \hat{n}_{m^{\prime} \sigma,0} \\ \nonumber
& -J \sum_{m \neq m^{\prime}} c_{m \uparrow,0}^{\dagger} c_{m \downarrow,0} c_{m^{\prime} \downarrow,0}^{\dagger} c_{m^{\prime} \uparrow,0}+J \sum_{m \neq m^{\prime}} c_{m \uparrow,0}^{\dagger} c_{m \downarrow,0}^{\dagger} c_{m^{\prime} \downarrow,0} c_{m^{\prime} \uparrow,0} \\ \nonumber
\hat{H}_\mathrm{bath} = &\sum_{\alpha,i=1}^{N_b}\epsilon^i_{\alpha}\hat{c}^\dagger_{\alpha,i}\hat{c}_{\alpha,i} + \sum_{\kappa,\alpha,i=1}^{N_b-1} 
       V^i_{\alpha_1\alpha_2}\hat{c}^\dagger_{\alpha_1,i} \hat{c}_{\alpha_2,i+1} + \text{h.c.}, \\ \nonumber
\hat{H}_\mathrm{hyb} = &\sum_{\kappa,\alpha}
    V^0_{\alpha_1\alpha_2}\hat{c}^\dagger_{\alpha_1,0} \hat{c}_{\alpha_2,1} + \text{h.c.}, 
\end{align}
where $\hat{H}_\mathrm{loc}$ represents the local interactions of the impurity, $\hat{H}_\mathrm{bath}$ denotes the bath, and $\hat{H}_\mathrm{hyb}$ describes the hybridization between the impurity and the bath. Here, $\alpha=\left(m,\sigma \right)$, with $m\in[1,N_o]$ and $\sigma=\left\{\uparrow,\downarrow\right\}$, denotes the orbital and spin degrees of freedom; $i$ is the bath site index, with $i=0$ representing the impurity, and $N_b$ is the number of bath sites per spin-orbital. Here, the bath takes the form of a chain with nearest neighbor hoppings.

In the main text, we consider the one-body energy and hybridization tensors to be diagonal in spin-orbital. The impurity on-site energies $\varepsilon_{\alpha_1\alpha_2}=\mu\delta_{\alpha_1\alpha_2}$ are varied to change its occupation. The bath on-site energies are all zero ($\epsilon^i_\alpha=0$), with nearest neighbor hoppings $V^{i}_{\alpha_1\alpha_2} = t\delta_{\alpha_1\alpha_2}$. This means each impurity spin-orbital is equivalently coupled to a bath with a semi-elliptic density of states with a half bandwidth of $D = 2t$.

\section*{S2 --- Optimization of NQS}
After parameterizing the many-body wave function using the ViT architecture as $\ket{\psi_{\bm{\theta}}} = \sum_{s} \psi_{\bm{\theta}}(s) \ket{s}$, the expectation value of an arbitrary operator $\hat O$ for a given set of parameters $\bm{\theta}$ is expressed as
\begin{equation}
\begin{split}
    O_{\bm{\theta}} &= \langle \hat{O} \rangle_{\bm{\theta}} 
    =\frac{\langle \psi_{\bm{\theta}} | \hat{O} | \psi_{\bm{\theta}} \rangle}{\langle \psi_{\bm{\theta}} | \psi_{\bm{\theta}} \rangle} \\ 
    &= \sum_{s} P_{\bm{\theta}}(s) O_{\bm{\theta}}^{\text{loc}}(s),
\end{split}    
\end{equation}
where $P_{\bm{\theta}}(s) = |\langle \psi_{\bm{\theta}} | s \rangle|^2 / \langle \psi_{\bm{\theta}} | \psi_{\bm{\theta}} \rangle$ represents the probability of configuration $\ket{s}$, and the corresponding local estimator
\begin{align}
    O_{\bm{\theta}}^{\text{loc}}(s) = \sum_{s'} \langle s | \hat{O} | s' \rangle \frac{\langle s' | \psi_{\bm{\theta}} \rangle}{\langle s | \psi_{\bm{\theta}} \rangle}.
\end{align}
For a given Hamiltonian $\hat{H}$, its energy $E_{\bm{\theta}}=\langle\hat{H}\rangle_{\bm{\theta}}$ is bounded from below by the exact ground state energy, which serves as a natural loss function for the optimization process. The variational parameters, $\bm{\theta}$, are iteratively optimized by minimizing this energy to approximate the ground state, typically employing a gradient-based approach controlled by a learning rate $\eta$:
\begin{equation}\label{eq:step}
    \theta'_k = \theta_k - \eta \left[S^{-1}\right]_{kk'} f_{k'},
\end{equation}
where the stochastic reconfiguration (SR) scheme~\cite{VMC_SR,VMC_McMillan,Giuseppe2017} is employed to account for the geometry of the learning landscape. It weights the force vector $f$ by the inverse of the quantum Fisher matrix $S$, which are defined as
\begin{equation}\label{eq:vmc_sr}
\begin{split}
    S_{kk'} &= \langle \mathcal{O}_{k}^{*} \mathcal{O}_{k'} \rangle_{\bm{\theta}} - \langle \mathcal{O}_k^{*} \rangle_{\bm{\theta}}  \langle \mathcal{O}_{k'} \rangle_{\bm{\theta}} , \\ 
    f_{k'} &=\langle \mathcal{O}_{k'}^{*} \left( E_{\bm{\theta}}^{\text{loc}} - E_{\bm{\theta}} \right) \rangle_{\bm{\theta}}, 
\end{split}
\end{equation}
where $\mathcal{O}_k(s) = \partial_{\bm{\theta}_k} \log\psi_{\bm{\theta}}(s)$.

For large systems, the dimension of the basis set $\{\ket{s}\}$ grows exponentially with system size $L$. Consequently, the exact evaluation of the expectation values $\langle \cdot \rangle_{\bm{\theta}}$ becomes computationally infeasible. Instead, these values are normally estimated using Monte Carlo importance sampling, and the summation is carried out using the approximate probability distribution over a much smaller number of states, $N_s$, generated by Monte Carlo. 

As discussed in the main text and also in~\cite{Lu2019,cao_tree_imp_2021}, in the natural orbital basis, QIMs exhibit an exponential decrease in electron (hole) leakage into the respective conduction (valence) bath chains as one progresses deeper into these chains. This behavior suggests that the ground state predominantly features density fluctuations at the impurity and its immediate neighboring sites, while deeper sites in the bath chains remain largely inactive. Consequently, the resulting probability distribution $P_{\bm{\theta}}(s)$ is expected to be highly skewed, with a few states contributing significantly, whereas the majority offer negligible weights. This skewed distribution profile impedes efficient sampling using traditional Monte Carlo methods.

Therefore, instead of using standard VMC methods to optimize the ViT NQS, we propose a subspace expansion method in the main text that leverages the skewed probability distribution to significantly improve the efficiency and accuracy of the estimation of expectation values. Within a given $\mathcal{S}$, similar to the Monte Carlo optimization, the expectation values in Eq.~\eqref{eq:vmc_sr} takes their approximate form
\begin{equation}
    \langle \hat{O} \rangle_{\bm{\theta}} \approx \sum_{s\in\mathcal{S}} \mathcal{P}_{\bm{\theta}}(s) O_{\bm{\theta}}^{\text{loc}}(s),
\end{equation}
where $\mathcal{P}_{\bm{\theta}}(s) = \sum_{s\in\mathcal{S}}\left( \frac{|\langle s | \psi_{\bm{\theta}} \rangle|^2}{\sum_{s'\in\mathcal{S}} |\langle s' | \psi_{\bm{\theta}} \rangle|^2} \right)$. The quantities for updating $\theta$ in Eq.~\eqref{eq:step} are evaluated as
\begin{align}\label{eq:se_sr}
   S_{kk'} &= \sum_{s\in\mathcal{S}} \mathcal{P}_{\bm{\theta}}(s) \mathcal{O}_{k}^{*}(s) \mathcal{O}_{k'}(s)   - \left(\sum_{s\in\mathcal{S}} \mathcal{P}_{\bm{\theta}}(s) \mathcal{O}_k^{*}(s) \right)  \left(\sum_{s\in\mathcal{S}} \mathcal{P}_{\bm{\theta}}(s) \mathcal{O}_{k'}(s) \right) , \\ \nonumber
   f_{k'} &= \sum_{s\in\mathcal{S}} \mathcal{P}_{\bm{\theta}}(s) \mathcal{O}_{k'}^{*}(s) \left( E_{\bm{\theta}}^{\text{loc}}(s) - E_{\bm{\theta}} \right), \\ \nonumber
   E_{\bm{\theta}} &= \sum_{s\in\mathcal{S}} \mathcal{P}_{\bm{\theta}}(s) E_{\bm{\theta}}^{\text{loc}}(s).
\end{align}

\section*{S3 --- $U$ and $N_b$ dependence}
\begin{figure}[t]
\centering
  \includegraphics[width=0.45\linewidth]{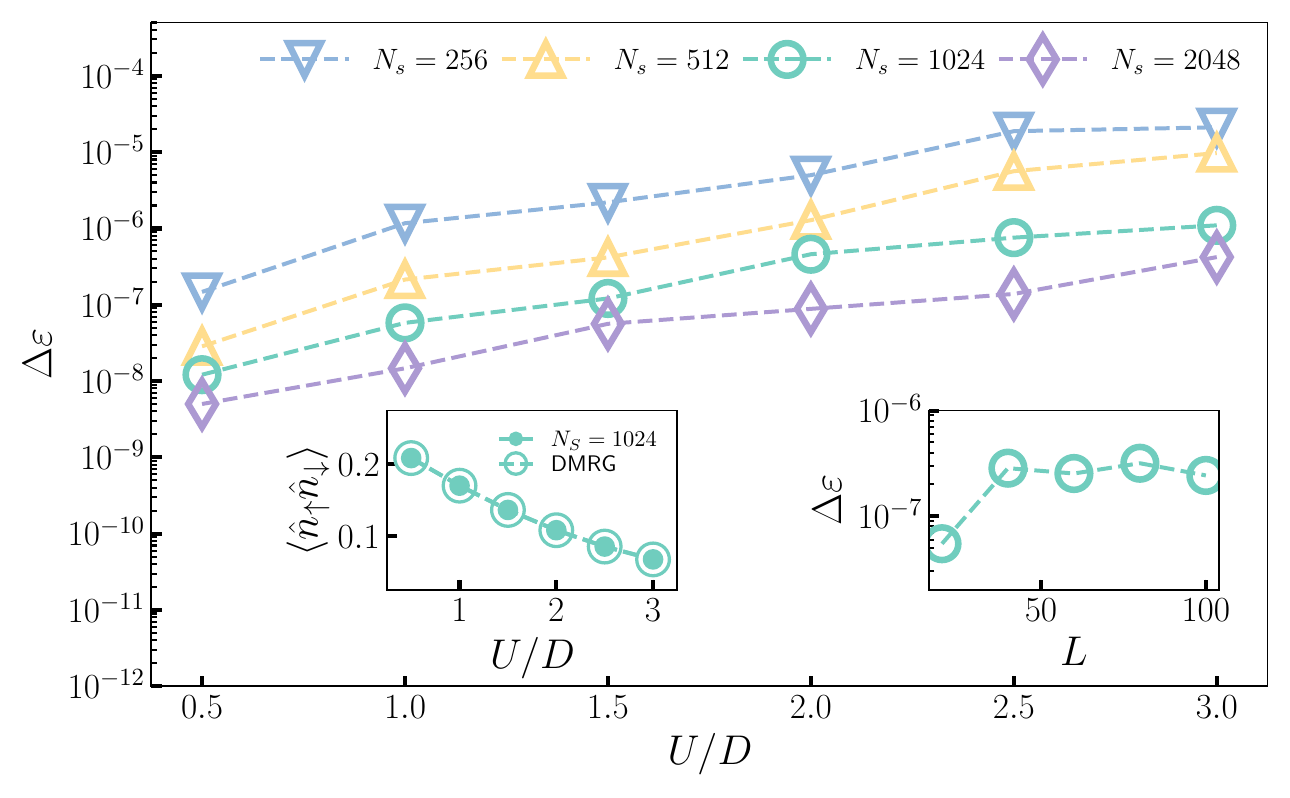}
  \caption{\label{fig::Udependence} Relative error $\Delta\varepsilon=\left(E_{\text{ViT}} - E_{\text{DMRG}}\right)/|E_{\text{DMRG}}|$ of the single-orbital Anderson impurity model with the NQS optimized by subspace expansion for various $N_s$ as a function of $U$. The left inset compares the double occupancy on the impurity site from ViT and MPS-DMRG. The right inset shows the relative error $\Delta\varepsilon$ as a function of system size. We use a ViT with $N_l=1$ and $d=32$. }
\end{figure}

To further illustrate the efficacy of our method, we evaluate the relative error of the ViT, configured with fixed $N_l=1$ and $d=32$, for different values of interaction strengths $U$ for the single-orbital Anderson impurity model at half-filling, as shown in Fig.~\ref{fig::Udependence}. The results show good agreement with MPS-DMRG results across a wide range of $U$ values. Notably, achieving comparable accuracy at higher $U$ values requires a moderate increase in the number of states $N_s$. Additionally, as depicted in the left inset of Fig.~\ref{fig::Udependence}, the double occupancy on the impurity site aligns closely with the MPS-DMRG results, further confirming the validity of our approach. More importantly, as depicted in the right inset of Fig.~\ref{fig::Udependence}, the accuracy of ViT remains nearly constant across different system sizes. Using the same ViT architecture, the efficient representation in the natural orbital basis allows the ViT to maintain comparable accuracy in systems up to $L=100$, even when maintaining a small subspace size ($N_s=1024$).

\section*{S4 --- Construction of restricted excitation space and evaluation of XAS}

\begin{figure}[t]
\centering
  \includegraphics[width=0.45\linewidth]{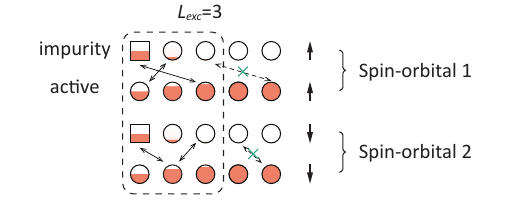}
  \caption{\label{fig::excitationIllu} Illustration of the construction of the XAS restricted excitation space for $N_o=1$. Only single-particle excitations within the cluster $\mathcal{C}_{L_\mathrm{exc}}$ are allowed.}
\end{figure}

As illustrated in Fig.~\ref{fig::excitationIllu}, the restricted excitation space $\{\ket{\phi_{\alpha,m}} \}$ introduced in the main text is defined by all possible single-particle excitations created during the XAS process within a cluster $\mathcal{C}_{L_{\text{exc}}}$ around the impurity site with size $L_{\text{exc}}$. Starting from the initial state
\begin{equation}
    \ket{\phi_{\alpha, m=0}} = D_\alpha \ket{\Psi_{\bm{\theta}}^{\text{GS}}} = \hat{c}^\dag_{\alpha,0} \ket{\psi_{\bm{\theta}}^{\text{GS}}},
\end{equation}
subsequent states are generated by applying all possible excitation operators $\hat T_{\beta, l, l'} = \hat{c}^\dagger_{\beta,l}\hat{c}_{\beta,l'}$ (where $l,l'\in \mathcal{C}_{L_{\text{exc}}},\ \beta\in [1,2N_o]$):
\begin{equation}
    \ket{\phi_{\alpha,m\geq 1}} = \hat{T}_{\beta, l, l'} \ket{\phi_{\alpha, m=0}}.
\end{equation}

As these states are not orthonormal, Eq.~\eqref{eq:xas} is then solved in the generalized form~\cite{noorthibasis}
\begin{equation}
    I^{\text{XAS}}_{\alpha}(\omega) = -\frac{1}{\pi}\mathrm{Im}\left[\mathbf{O} \left[\left(\omega +E_0 + i\Gamma\right) \mathbf{O} - \mathbf{H} \right]^{-1} \mathbf{O} \right]
\end{equation}
where
\begin{equation}
    \begin{split}
        \mathbf{O}_{m,m'} &= \langle \phi_{\alpha,m} | \phi_{\alpha,m'} \rangle, \\
        \mathbf{H}_{m,m'} &= \langle \phi_{\alpha,m} | \hat{H}_c |\phi_{\alpha,m'} \rangle.
    \end{split}
\end{equation}

\end{document}